\documentclass{epl}
\usepackage{graphicx}
\usepackage{bm}
\begin{document} 

\title{Maximum-likelihood absorption tomography}
\shorttitle{Maximum-likelihood absorption tomography}
\author{J. \v{R}eh\'{a}\v{c}ek\inst{1}
\thanks{Email: \email{rehacek@phoenix.inf.upol.cz}}
\and Z. Hradil\inst{1} \and
M. Zawisky\inst{2} 
\and W. Treimer\inst{3} \and 
M. Strobl\inst{3}}  
\shortauthor{J. \v{R}eh\'{a}\v{c}ek {\it et al.}}
\institute{
\inst{1} Department of Optics, Palack\'{y} University,
17. listopadu 50, 772 00 Olomouc, Czech Republic\\
\inst{2} Atominstitut der \"{O}sterreichischen Universit\"{a}ten, 
Stadionallee 2, A-1020 Wien\\
\inst{3} University of Applied Sciences (TFH) Berlin, FB II, 
Luxemburgerstra{\ss}e 10, D-13353 Berlin and Hahn-Meitner-Institute 
Berlin, Glienicker Stra{\ss}e 100, D-14109 Berlin
}


\maketitle

\begin{abstract}
Maximum-likelihood methods are applied to the problem 
of absorption tomography. The reconstruction is done 
with the help of an iterative algorithm. We show how the statistics
of the illuminating beam can be incorporated
into the reconstruction.  
The proposed reconstruction method can be considered as a useful 
alternative in the extreme cases where the standard ill-posed
direct-inversion methods fail.
\end{abstract}

\section{Introduction}

The standard reconstruction method in present computerized 
tomographic (CT) imaging is the filtered back-projection  (FBP)
algorithm which is based on the Radon transformation \cite{kak87}. 
Unfortunately FBP fails in case of  missing projections and/or if strong 
statistical fluctuations of the counting numbers are present in the small 
detector pixels. The latter situation occurs e.g. in neutron tomography 
\cite{schillinger99,schillinger00,koerner01,mcmahon01}, if monochromatic 
neutron beams are applied in order to avoid beam artifacts 
\cite{dubus02} or at the investigation of strong absorbing 
materials. 
The cases of missing projections and incomplete data sets  for
monochromatic neutron beams have been already investigated in the past in
detail by means of algebraic reconstruction technique 
\cite{treimer91,maas92,treimer98}.
Scattering data from a double crystal diffractometer have been used to
reconstruct 2D scattering pattern and the results were compared with the
standard FBP. With this algebraic approach one could reconstruct 
2D pattern in spite of the lack of nearly $90$ degrees of the scanning angle, 
whereas in such cases the FBP method entirely failed. The computing time, 
however, was extremely long (up to several hours), so that this method is 
useful for rather small 2D arrays ($100\times 100$  pixels) only.

The new reconstruction method proposed in this paper can improve several 
tomographic applications in neutron optics which in many cases 
are limited by the weak intensity and the poor detector resolution. 
The use of well collimated pencil beams which are scanned across the 
sample surface could dramatically enhance the spatial image 
resolution but this method is only rarely used due the long 
measurement times \cite{allman00}. An improved reconstruction method can 
encourage new applications in neutron optics which often suffer from the low    
counting numbers. Generally the new algorithm can achieve 
better reconstruction results or reduce the scanning time 
in neutron optics and in medical and biological CT imaging.

\section{LinPos tomography} \label{sec-linpos}

\begin{figure}
\twofigures[width=6cm]{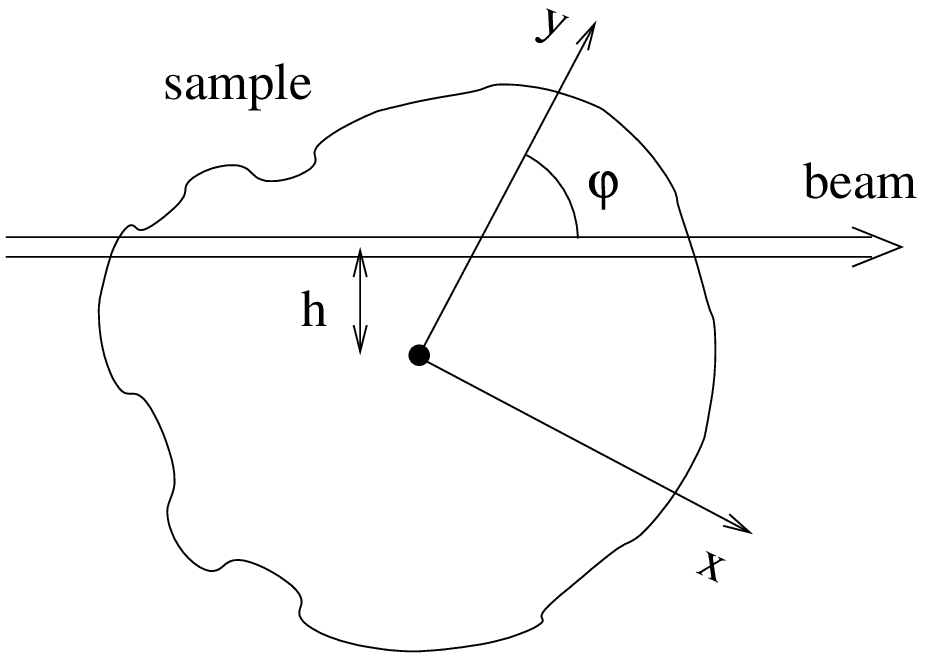}{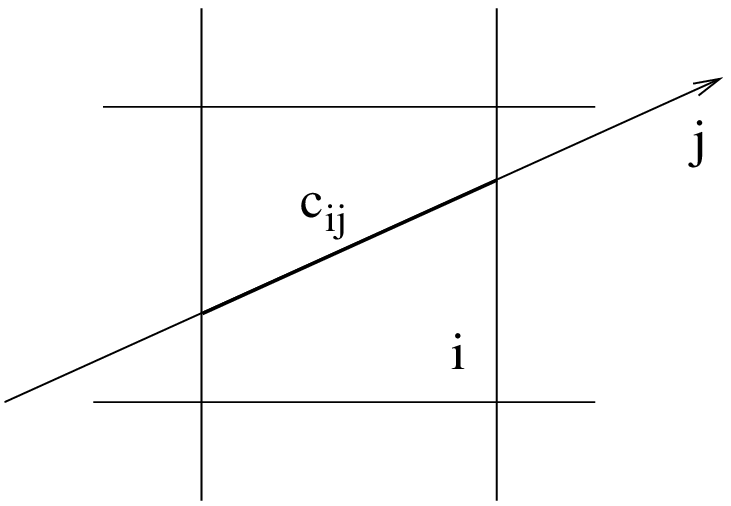}
\caption{Geometry of the experimental setup}
\label{fig-geom}
\caption{Definition of coefficients \(c_{ij}\)}
\label{fig-square}
\end{figure}

Basic notions and the geometry of experimental setup are
as follows. Let us assume that the sample is illuminated
by parallel monochromatic pencil beams, see Fig.~\ref{fig-geom}.
Data consist of the number of particles counted behind the sample
for $M$ different scans -- each scan being characterized by
horizontal position $h$ and rotation angle $\varphi$.
Alternatively, a broad illuminating beam combined with  a 
position-sensitive detector (CCD camera) placed behind the sample
can be used. In that case $h$ labels pixels of the camera.
For the sake of simplicity a collective index  $j\equiv\{h,\varphi\}$ 
will be used, hereafter, to label the scans.

Mean number $\bar{n}_j$ of particles (intensity) 
registered in $j$-th scan is given by the exponential law
%%%%%%%%%%%%%%%%%%%%%%%%%%%%%%%%%%%%%%%%%%%%%%%%%%%%%%%%%%
\begin{equation} \label{expon}
\bar{n}_j=\bar{n}_0 \exp(-\int \mu(x,y)ds_j),
\end{equation}
%%%%%%%%%%%%%%%%%%%%%%%%%%%%%%%%%%%%%%%%%%%%%%%%%%%%%%%%%%%
where $\bar{n}_0$ is the intensity of the incoming beam, $\mu(x,y)$
is the absorption index (cross section) of the sample in position
$\{x,y\}$, and the integration is the path integration along the 
pencil beam. This exponential attenuation law is a good approximation 
if scattering can be neglected. The beam hardening artifacts 
would also modify Eq. (\ref{expon}) but this complication can be 
avoided experimentally by the use of monochromatic beams 
\cite{dubus02}.
For practical purposes, it is convenient to discretize 
Eq.~(\ref{expon}) as follows,
%%%%%%%%%%%%%%%%%%%%%%%%%%%%%%%%%%%%%%%%%%%%%%%%%%%%%%%%%%%
\begin{equation} \label{discrete}
\bar{n}_j=\bar{n}_0 \exp(-\sum\limits_{i=0}^N \mu_i c_{ij}).
\end{equation}
%%%%%%%%%%%%%%%%%%%%%%%%%%%%%%%%%%%%%%%%%%%%%%%%%%%%%%%%%%% 
The sample is now represented by a 2D mesh. Each cell is assumed
to have a constant absorption index. The variables are now 
$N$ numbers $\mu_i$ specifying absorption
indices of those cells. Overlaps between beams and cells  
are stored in the array $\{c_{ij}\}$, see Fig.~\ref{fig-square}.

Let us first ignore the statistics of the illuminating beam,
and assume that the counted numbers of particles $\{n_j\}$ do not 
fluctuate, $n_j=\bar{n}_j,\;\forall j$.
Taking logarithms of both sides of Eq.~(\ref{discrete}), one obtains a 
system of $M$ linear algebraic equations for $N$ unknown 
absorption coefficients $\mu_i$:
%%%%%%%%%%%%%%%%%%%%%%%%%%%%%%%%%%%%%%%%%%%%%%%%%%%%%%%%%%%
\begin{equation} \label{linpos}
f_j=p_j,\quad j=1\ldots M,
\end{equation}
%%%%%%%%%%%%%%%%%%%%%%%%%%%%%%%%%%%%%%%%%%%%%%%%%%%%%%%%%%%
where we defined,
\begin{equation} \label{def}
f_j=-\ln\frac{n_j}{n_0}, \quad
p_j=\sum_i\mu_i c_{ij}.
\end{equation}
Notice that problem (\ref{linpos}) is a linear and 
positive (LinPos) problem. Positivity follows from the fact that
no new particles are created in the sample.
Although direct inversion of Eq.~(\ref{linpos}) is possible 
for $N\geq M$, the solution is not always positively defined.
A negative value of a reconstructed $\mu_i$ would suggest 
that particles were being created in the $i$-th cell
in the course of the experiment, which would obviously be
a wrong conjecture. This problem can be avoided if the
problem (\ref{linpos}) is solved in the sense of maximum likelihood
(ML) on the  space of physically allowed absorption coefficients. 
In this approach one considers the data $\bm{f}$ and the 
prediction of the theory $\bm{p}$ as two probability
distributions.  
One looks for absorption coefficients $\{\mu_i\}$ 
that minimize the Kullback-Leibler ``distance'' 
\begin{equation} \label{kullback}
d(\bm{f},\bm{p})=-\sum\limits_j f_j\ln \frac{p_j}{f_j}
\end{equation}
between the data $\bm{f}$ and the theory $\bm{p}$. 
Here a little extra care is
needed since $\bm{p}$ and $\bm{f}$ are generally not
normalized to unity. The minimum of the Kullback-Leibler
distance corresponds to the maximum of the maximum likelihood (ML) 
functional \cite{kendall61}
%%%%%%%%%%%%%%%%%%%%%%%%%%%%%%%%%%%%%%%%%%%%%%%%%%%%%%%%%%%%
\begin{equation} \label{lik}
{\cal L}=\prod\limits_j \left(\frac{p_j}{\sum_k p_k}\right)^{f_j},
\end{equation}
%%%%%%%%%%%%%%%%%%%%%%%%%%%%%%%%%%%%%%%%%%%%%%%%%%%%%%%%%%%%
that quantifies the likelihood of the given distribution
$\{\mu_i\}$ in view of the registered data. We seek the
maximum-likely distribution of the absorption indices. 
A convenient way how to find it is the celebrated  
Expectation-Maximization (EM) iterative algorithm
\cite{dempster77,vardi93},
%%%%%%%%%%%%%%%%%%%%%%%%%%%%%%%%%%%%%%%%%%%%%%%%%%%%%%%%%%%%
\begin{equation} \label{zdenek}
\bm{\mu}^{n+1}=\bm{R}(\bm{\mu}^n)\cdot\bm{\mu}^n,
\end{equation}
%%%%%%%%%%%%%%%%%%%%%%%%%%%%%%%%%%%%%%%%%%%%%%%%%%%%%%%%%%%%
where 
%%%%%%%%%%%%%%%%%%%%%%%%%%%%%%%%%%%%%%%%%%%%%%%%%%%%%%%%%%%%
\begin{equation} \label{r}
R_i=\frac{1}{\sum\limits_{j'} c_{ij'}}\sum\limits_j
\frac{f_j c_{ij}}{p_j(\bm{\mu})},
\end{equation}
%%%%%%%%%%%%%%%%%%%%%%%%%%%%%%%%%%%%%%%%%%%%%%%%%%%%%%%%%%%%
and $\bm{\mu}^0$ is some initial strictly positive distribution 
$\mu_i^{(0)}>0,\, i=1\ldots N$. A nice feature of EM algorithm is that
its convergence is guaranteed for any input data $f_j$ \cite{shepp82}. 
For this reason it became a valuable tool in many inverse problems which 
can be reduced to the form of Eq.~(\ref{linpos}), e.g. in positron 
emission tomography \cite{shepp82,vardi85,mair96}.
The original derivation of EM algorithm is based on alternating
projections on specially chosen convex sets of vectors.
However, one could directly use the calculus of variations
to derive the necessary condition for the extreme of the functional 
(\ref{lik}). Iterating these, one eventually arrives at the 
EM algorithm again. An advantage of this alternative derivation
is that it can be also applied to more realistic physical 
models of the actual absorption experiment. One such possible
generalization will be studied in the following section.

\section{Tomography with Poissonian signals} \label{sec-poiss}

Real signals are not composed of a sharp number of particles.
For instance, two signals often used in experiments
---beam of thermal neutrons and laser light--- both exhibit
Poissonian fluctuations in the number of particles. 
Also monochromatic neutron beams are correctly described by 
Poissonian statistics if the detected count events occur 
mutually independently \cite{rauch90}.
The knowledge of the true character of signal illuminating the sample is a 
useful piece of prior information, which can be utilized
for improving the performance of ML tomography.

As the Poissonian character of the signal is 
preserved by the process of attenuation, the counted numbers
of particles behind the sample are random Poissonian
variables. The corresponding likelihood functional
reads,
%%%%%%%%%%%%%%%%%%%%%%%%%%%%%%%%%%%%%%%%%%%%%%%%
\begin{equation} \label{poiss-lik}
{\cal L}\propto \prod\limits_j \bar{n}_j^{n_j}e^{-\bar{n}_j}.
\end{equation}
%%%%%%%%%%%%%%%%%%%%%%%%%%%%%%%%%%%%%%%%%%%%%%%%  
This is the joint probability of counting $\{n_j\}$
particles. Mean values $\{\bar{n}_j\}$ obey the exponential
law (\ref{expon}) as before. They depend on the absorption
in the sample $\{\mu_j\}$ that is to be inferred from the data.
The necessary condition for the extreme of the likelihood
(\ref{poiss-lik}) can be derived using the calculus of variations.
The extremal equation can be shown to have the same vector form 
as the extremal equation of the LinPos problem 
(\ref{zdenek}). The vector $\mathbf{R}$ now becomes
%%%%%%%%%%%%%%%%%%%%%%%%%%%%%%%%%%%%%%%%%%%%%%%%
\begin{equation} \label{r-poiss}
R^{\mathrm{(Poisson)}}_i=\frac{\bar{n}_0}{\sum\limits_{j'}
c_{ij'}n_{j'}}\sum\limits_{j}c_{ij}\exp (-\sum\limits_{i'}
\mu_{i'}c_{i'j}).
\end{equation}
%%%%%%%%%%%%%%%%%%%%%%%%%%%%%%%%%%%%%%%%%%%%%%%%
When the input intensity $\bar{n}_0$ is not known,
it can be estimated together with the absorption of
the sample:
%%%%%%%%%%%%%%%%%%%%%%%%%%%%%%%%%%%%%%%%%%%%%%%%%
\begin{equation} \label{n0}
\bar{n}_0=\frac{\sum\limits_j n_j}{\sum\limits_j
\exp(-\sum\limits_i \mu_ic_{ij})}.
\end{equation}
%%%%%%%%%%%%%%%%%%%%%%%%%%%%%%%%%%%%%%%%%%%%%%%%%%
Poissonian tomography is intrinsically a nonlinear
problem. This has serious consequences for
the convergence properties of the iterative algorithm (\ref{zdenek}) and 
(\ref{r-poiss}).  Instead of  converging to a stationary point it might 
end up in  oscillations. Typically such convergence problems 
arise in the presence of very noisy data.
When this happens one can always decrease the length of the iteration step 
as follows:
$R_i \rightarrow R_i^{\alpha},\quad i=1\ldots M,\quad 0<\alpha<1$.
Of course, any solution to the regularized problem
is also  a solution to the original problem.

\section{Discussion}

Generally, the reconstructed image will depend on which ML method
is chosen to process the data; see the apparent difference 
between  Eqs.~(\ref{r}) and (\ref{r-poiss}).
It is interesting to look more closely at the origin of this
difference. Consider a tomographic setup with a Poissonian beam.
Then the Poissonian algorithm should provide a better 
reconstruction than the LinPos algorithm which have been 
derived under the assumption of non-fluctuating signals. 
The LinPos reconstruction consists in minimizing the Kullback-Leibler 
distance between the data $\bm{f}$ and theory $\bm{p}$. 
When logarithms of the counted numbers of particles
are chosen to be the input data rather than counted data itself, 
one arrives at the EM algorithm (\ref{zdenek}) and (\ref{r}). 
Taking logarithms of actual data makes the problem linear and 
considerably simplifies the reconstruction.
However, one could, instead, directly minimize the Kullback-Leibler
distance between the counted data $n_j$ and the corresponding theory 
$p'_j=n_0\exp(-p_j)$.  Interestingly enough, the extremal equations 
associated with this variational problem are the same as
Eqs.~(\ref{zdenek}) and (\ref{r-poiss}) derived above from the Poissonian 
theory (\ref{poiss-lik}). Choosing $n_j$ instead of $f_j$ as the data is 
equivalent to taking the Poissonian statistics of the signal into account!
The difference between the LinPos and Poissonian ML reconstructions
can thus be traced down to whether the measured data
are used directly or not. Tampering with data
prior to reconstruction may speed up and facilitate
the whole process of reconstruction but some information 
about the object might get lost. 

\section{Comparison with standard methods}

\begin{figure}
\onefigure[angle=270,width=3.5cm]{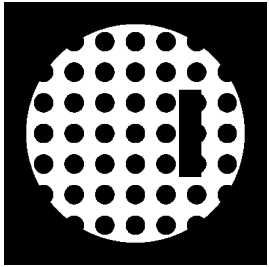}
\caption{The object.}
\label{fig:object}
\end{figure}

In a real experiment there are many factors that could 
influence the quality of the measured data and therefore 
also on the result of the tomography.
Misalignments present in the experimental setup, instability of the 
illuminating beam, white spots and damaged detector pixels can be such 
factors, to name a few. To avoid this problem we replaced the experiment
by a simulation. The data were generated on a computer. The artificial 
object used in the simulation is shown in Fig.~\ref{fig:object}. 
The object is a circle made of a homogeneous material with many small 
round holes drilled through it. One additional rectangular
piece of material was removed from the circle to make it less 
symmetric. Absorption index of the material was chosen in such a way
that the maximum attenuation along a beam was close to $50\%$ of the 
input intensity.
 
In the simulation, the object was subject 
to five different experiments. Their parameters are summarized
in Table~\ref{tab:recon-param}.
\begin{table}
\begin{tabular}{cccc}
reconstruction & angles & pixels & intensity\\
\hline
a & 13 & 161 & $\infty$\\
b & 19 & 101 & $\infty$\\
c & 20 & 101 & $\infty$\\
d &  7 & 301 & $\infty$\\
e & 15 & 161 & 2000 
\end{tabular}
\caption{Quality of the input data. The last column
shows the mean number of counted particles per pixel 
in the incident beam.}
\label{tab:recon-param}
\end{table}
First four experiments correspond to the ideal situation of a 
very high beam intensity where the Poissonian detection noise
can safely be ignored. The last reconstruction simulates more 
realistic conditions with $2000$ counts per pixel in the open beam.
Notice that a relatively small number of rotations is chosen for all 
five experiments. In this regime the Radon transformation is expected
to yield bad results and the improvement of the maximum-likelihood 
tomography upon the standard technique should be most prominent.
This regime is also important from the practical point of view.
Doing more rotations implies a longer measurement time and more radiation
absorbed by a sample. The latter may be an important factor if the 
imaging of biological samples is considered. So, imaging costs
and damage done to a sample due to radiation might be reduced 
provided the improvement of the reconstruction technique gives comparable
resolution with less data. 

\begin{figure}
\twofigures[width=6cm]{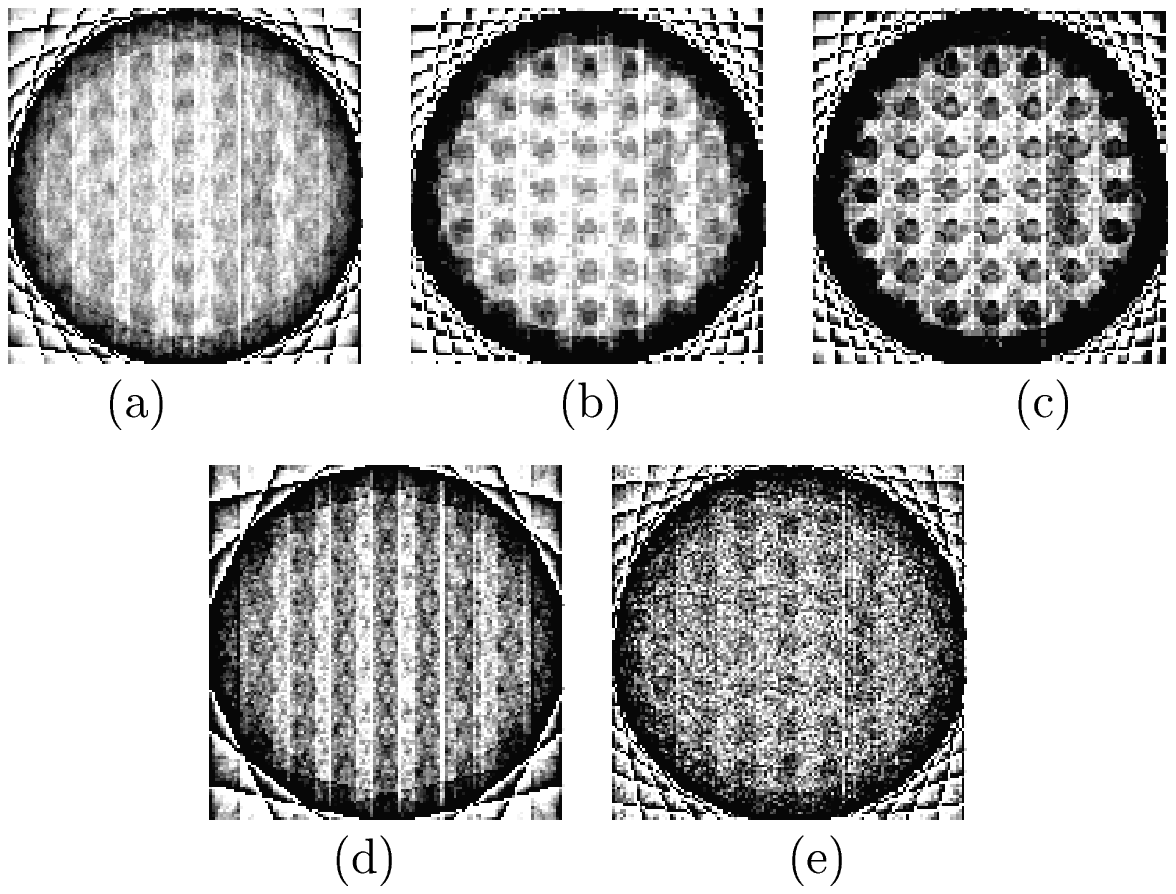}{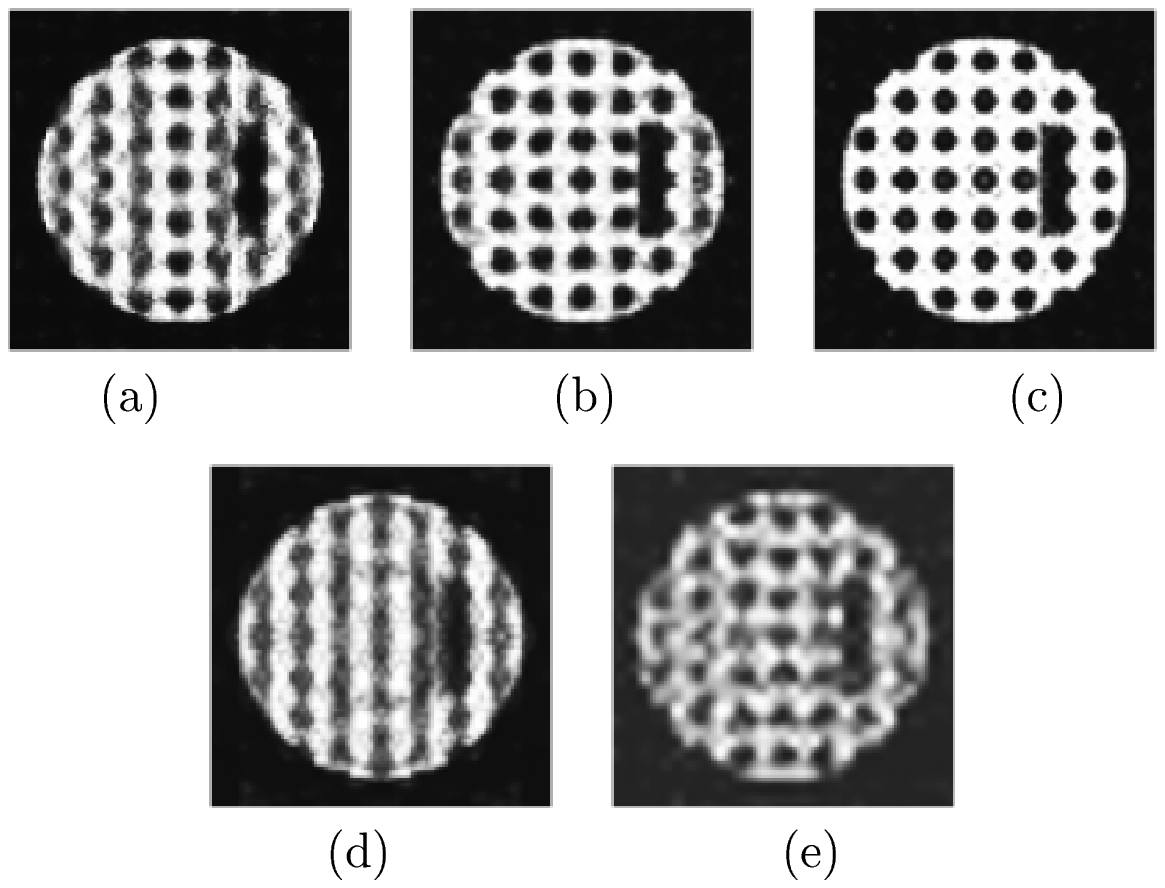}
\caption{IDL reconstructions from the simulated data,
for parameters see Tab.~\ref{tab:recon-param}}
\label{fig:IDL}
\caption{ML reconstructions from the same data.
The proposed iterative  algorithm, Eqs.~(\ref{zdenek}) and 
(\ref{r-poiss}), has been used for reconstruction.} 
\label{fig:ML}
\end{figure}

Reconstructions from the simulated data are shown in 
Figs.~\ref{fig:IDL} and \ref{fig:ML}.
The simulated data were first processed using the IDL imaging 
software (Research Systems Inc.) which implements the standard FBP
algorithm  (Radon transform), see Fig.~\ref{fig:IDL}. This 
software is one of the industrial standards in the computer assisted 
tomography.  The same data were then processed using our iterative 
algorithm based  on the maximization of the Poissonian likelihood 
function, see Fig.~\ref{fig:ML}.
In the absence of noise, see cases (a)-(d), the fidelity of a 
reconstruction depends on two main factors---the spatial resolution of the
detector, and the number of rotations used. It is apparent from
Figs.~\ref{fig:IDL} and \ref{fig:ML} that the latter factor is more 
important of the two. Very small number of angles cannot be compensated
by an increased spatial resolution of the detector, compare e.g. 
cases (c) and (d),
and reconstruction (d) is by far the worst one. However, ML tomography
is much less sensitive to the number of angles than the standard 
filtered back-projection. Even the large rectangular
hole in the object is hardly perceptible in Fig.~\ref{fig:IDL}d 
whereas it nicely shows in the ML reconstruction
Fig.~\ref{fig:ML}d. ML reconstructions are superior to the standard ones
also in cases (a)-(c); notice that the reconstruction Fig.~\ref{fig:ML}c done 
with as few as $20$ different angles is nearly perfect. 

Benefits of the ML tomography are fully revealed when the detected
data are noisy. This is case (e) in  Tab.~\ref{tab:recon-param}. 
Standard filtered
back-projection applied to noisy data faces serious difficulties. This is 
due to ill-poseness of the Radon transformation where data are integrated
with a singular filter function. Obviously such deconvolution
greatly amplifies any noise present in the data. 
Having little or no prior information about the object it is difficult
to tell true details of the object from artifacts.
ML tomography gives much better results.  
Since noises are incorporated into the algorithm in a natural and 
statistically correct way artificial smoothing is not needed. 
Notice in Fig.~\ref{fig:ML}e that noisy data yield a little distorted but 
otherwise clear image unlike the corresponding very noisy standard 
reconstruction  shown in Fig.~\ref{fig:IDL}e. This is a nice feature of the 
intrinsically nonlinear ML algorithm which,  in the course of reconstruction, 
self-adapts to the registered data and always selects the most likely 
configuration.

Finally let us emphasize that apart from the size of the reconstruction mesh 
$N$ \cite{mesh} there are no free parameters left in the ML algorithm 
to play with. This prevents one from interfering when the reconstructed
image ``looks bad.'' This also makes the whole procedure more objective,
which is a necessary presumption for the investigation of ultimate
limits of reconstruction schemes.

\section{Conclusion}
We presented a new reconstruction method for CT imaging based
on the iterative maximization of the Poissonian likelihood.
For small number of scans and/or short measurement time this method 
was shown to yield a significant improvement upon the 
standard filtered back-projection algorithm. This could be important
for CT imaging with low-intensity beams, and for
applications where strong irradiation of a sample during the scanning
should be avoided. One area where reconstruction techniques
of the type discussed in this paper would be very useful are coherent
reconstruction techniques such as interferometric phase tomography
with X-rays \cite{momose95,beckmann97} or neutrons \cite{badurek00}, 
or neutron holography \cite{cser01}. There is hopefully more to come.

\acknowledgments
This work was partially supported by Grant No. LN00A015 of the Czech Ministry
of Education (J.\v{R} and Z.H.), by Austrian Science Foundation, 
project No P14229-PHY (M.Z.), by the BMBF, project 03TRE9B6 (W.T. and M.S), 
and by the TMR-Network of the European  Union 
``Perfect Crystal Neutron Optics,'' ERB-FMRX-CT96-0057.
\acknowledgments


\begin{thebibliography}{10}
\bibitem{mesh}
$N$ can always be reasonably chosen based on the size $M$ of the measured 
data. The rule of thumb is $N\approx M$.
\end{thebibliography}
\end{document}